\documentclass[aps,prl,preprint,superscriptaddress,amsmath,amssymb]{revtex4}

\usepackage{graphicx}
\usepackage{dcolumn}
\usepackage{bm}

\begin{document}

\title{Strong Electron-Hole Exchange in Coherently Coupled Quantum Dots}

\author{Stefan F\"alt}
\affiliation{Institute of Quantum Electronics, ETH Zurich, 8093 Zurich, Switzerland}
\author{Mete Atat\"ure}
\affiliation{Cavendish Laboratory, University of Cambridge, Cambridge, CB3 0HE, United Kingdom}
\author{Hakan Tureci}
\affiliation{Institute of Quantum Electronics, ETH Zurich, 8093 Zurich, Switzerland}
\author{Yong Zhao}
\affiliation{Physikalisches Institut, Ruprecht-Karls-Universit\"at Heidelberg, 
Philosophenweg 12, 69120 Heidelberg, Germany}
\author{Antonio Badolato}
\affiliation{Institute of Quantum Electronics, ETH Zurich, 8093 Zurich, Switzerland}
\author{Atac Imamo\u{g}lu}
\affiliation{Institute of Quantum Electronics, ETH Zurich, 8093 Zurich, Switzerland}

\date{\today}

\begin{abstract}
We have investigated few-body states in vertically stacked quantum dots. Due to 
small inter-dot tunneling rate, the coupling in our system is in a previously 
unexplored regime where electron-hole exchange is the dominant spin interaction. 
By tuning the gate bias, we are able to turn this coupling off and study a 
complementary regime where total electron spin is a good quantum number. The use 
of differential transmission allows us to obtain unambiguous signatures of the 
interplay between electron and hole spin interactions. Small tunnel coupling 
also enables us to demonstrate all-optical charge sensing, where conditional 
exciton energy shift in one dot identifies the charging state of the coupled 
partner.
\end{abstract}

\maketitle

Self-assembled quantum dots (QDs) are semiconductor nanostructures that exhibit 
three-dimensional confinement of carriers. Due to spatial confinement, the electronic 
states in a QD are quantized and these structures have been referred to as artificial 
atoms. This description has been demonstrated experimentally with atom-like properties 
such as strong photon antibunching \cite{P.Michler12222000} and near
life-time limited linewidths \cite{seidl:195339}. One can use the self-assembly 
mechanism to create aligned nanostructures that
function as QD molecules. Eearlier studies have demonstrated hybridization of energy 
levels of two coupled QDs \cite{krenner:057402}, spectral signatures of tunnel 
coupling of multiple-hole \cite{E.A.Stinaff02032006,scheibner:245318} or 
two-electron \cite{krenner:076403} states. It has also been shown that the g-factor can 
be engineered with control of 
the tunneling \cite{doty:197202}. In contrast to the previously well studied coupling 
regime \cite{krenner:076403}, in our system the inter-dot electron coupling mechanism
is significantly modified by electron-hole exchange. By tuning the gate bias, the 
complementary regime of pure electron-electron exchange was also investigated. Here, 
we use resonant scattering techniques to resolve the spectral signatures of various 
coupling mechanisms. We studied the optical emission and absorption from both QDs in 
the pair, which allowed us to use inter-dot Coulomb interactions in the system to 
determine the charging state of both QDs.

The heterostructures for the device for this work was grown with
molecular beam epitaxy on a GaAs substrate. The sample consists of
two layers of QDs, separated with 15 nm of GaAs, in a diode
structure. The back contact consists of a n-doped layer and the top
contact is a semi-transparent layer of Ti. The QD layer close to the
back contact was separated from it with 25 nm GaAs and more
blue-shifted than the layer closer to the top contact. The layers
will be referred to as the blue and red layers, respectively. The
red layer was spaced from the top gate with 160 nm GaAs, including
an AlGaAs current blocking layer close to the top gate. The strain
field on top of QDs from the blue layer gives a natural alignment
of the nucleation of QDs in the red layer so that stacks are formed.

The measurements were performed using micro-photoluminescence
($\mu$-PL) and differential transmission (DT) techniques at 4.2 K.
For $\mu$-PL, a 780 nm laser was used to create free carriers in the
bulk GaAs. A lens with NA of 0.55 was used to both focus this laser
and collect the luminescence, which was spectrally resolved with a
resolution of 30 $\mu$eV. For DT measurements, a single frequency
laser was tuned across the coupled QD resonances. A Si p-i-n
photodiode detected the transmitted laser light and a lock-in
amplifier was used with stark-shift modulation of the resonances to
eliminate low frequency noise \cite{alen:2235}.

The diode structure allows controlled electron charging of the QDs.
As the stacking probability of the QDs is not unity, we were able use
single QDs without a coupled partner to investigate the spectral
signatures of charging in the absence of coupling. These QDs showed
no fundamental differences from what is known about single QDs in
charge-injection devices. The charging behavior changes with the
potential well depth, but also with the distance to the doped
layer. In addition to emission wavelength, this difference in
charging behavior between the layers aids us in the identification
of which layer a certain QD is in.

\begin{figure}
\includegraphics[width=82mm]{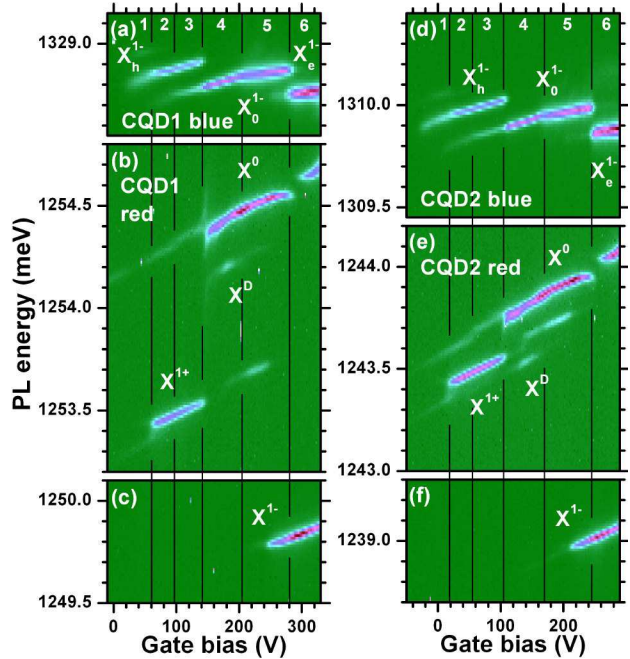}
\caption{\label{fig:PL} PL plots of two different pairs of stacked
QDs as a function of the applied gate bias where (a) and (d) show PL
data from the blue QDs of the pair, (b), (c), (e), and (f) show the data
from the red QDs. The PL lines are identified with the corresponding excitonic 
states. Regions with different charge combinations are separated with black 
vertical lines and numbered. Using the notation $(B,R)$, where B denotes the ground 
state charge of the blue QD and R that of the red QD, the ground state charge 
configurations are: 1: $(0,0)$ or $(0,h^+)$, 2: $(0,0)$ or $(e^-,h^+)$, 3: $(e^-,0)$ 
or $(e^-,h^+)$, 4: $(e^-,0)$, 5: $(2e^-,0)$ and 6: $(2e^-,e^-)$.}
\end{figure}

The relatively large tunnel barrier between our  dots allows us to
study the emission properties of each QD of a pair with $\mu$-PL. In
Figure 1, we present data from two pairs of coupled QDs (CQD1 and
CQD2), focusing on three separate PL energy windows in the same gate
voltage range. Figures 1 (a) and (d) show the PL intensity lines of
the negatively charged trion on the blue dot. Figures 1 (b) and (e)
contain the the lines for the neutral bright exciton (X$^{0}$), dark
exciton (X$^{D}$) and positively charged trion (X$^{1+}$), while
Figures 1 (c) and (f) show the negatively charged trion (X$^{1-}$)
lines \cite{Warburton:5761} for the red QDs. We first note that the
X$^{1-}$ line of the blue QDs in Figures 1 (a) and (d) is split into
three, following closely the charging state of their partners in
Figures 1 (b) and (e) respectively. In other words, the wavelength
of the X$^{1-}$ line of the blue QDs is conditional on the ground
state of its red partner. We indicate this using the following
notation: X$^{1-}$ for a single hole charge (X$^{1-}_h$), neutral
(X$^{1-}_0$) and single electron charge (X$^{1-}_e$) on the red dot.
The measured splittings between regions 5 and 6 (110 $\mu$eV) and 3
and 4 (130 $\mu$eV)  is consistent with the estimated dipole shifts
of the order of 100 $\mu$eV extracted from the dc Stark shift of the
lines. We note here that the shifts of the lines {\em in the red QD}
due to charge sensing in comparison is smaller and in the opposite
direction. While the latter observation is easily explained by the
charge being on the opposite side of the dipole as compared to the
case of the blue QD, the latter indicates a different spatial
composition of the red QD due to the strain field of their partners
in the first layer. This possibility is consistent with the X$^{1+}$
line being red-shifted as compared to the neutral exciton in
contrast to previous reports on single-dots \cite{ediger:211909}.

\begin{figure}
\includegraphics[width=82mm]{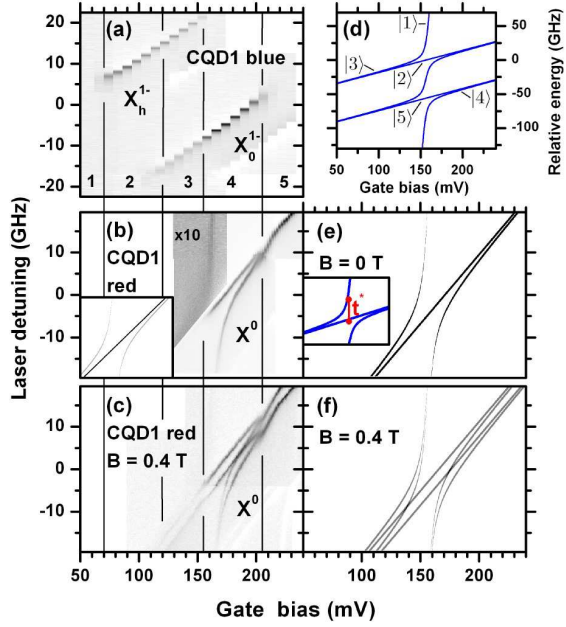}
\caption{\label{fig:DT} (a) DT measurements showing two split lines for the 
blue X$^1-$ due to charge sensing. Charges can be introduced into the red QD in 
this case due to excitation to higher excited state configurations. (b) 
DT measurements on the red X$^0$ and (c) red X$^0$ in an external magnetic field of 
0.4 T. (d) Calculated optically excited state level diagram as a function of gate 
bias. Calculated absorption (e) without and (f) with magnetic field. The inset to 
(b) shows the expected absorption for vanishing electron-hole exchange. The inset 
to (e) indicates the definition of the coupling strength $t^*$ in our system.}
\end{figure}

In order to investigate the spin fine structure in the regions 3 and
4, we now focus on the DT measurements which provide higher
resolution and eliminate spurious effects associated with the
generation of free charges in the bulk GaAs. Figures 2(a) and 2(b)
show data from the DT measurements carried out at zero external
magnetic field on the blue and red QD, respectively. The blue QD
shows a splitting due to the charge sensing that we noted earlier in
the PL data. At 205 mV gate bias when the blue dot charging changes, we see a kink 
in the red QD DT signal; at present, we do not understand this feature. Figure 2(c) 
shows the counterpart of Fig. 2(b) with an
applied magnetic field 0.4 T in Faraday configuration \footnote{The
splitting of the red X$^{0}$ line in Fig. 2(c) would correspond to
an exciton g-factor of only 0.6, which is the expected contribution
from the electron. Considering the unusual energy of X$^{+1}$
relative to X$^{0}$, we could argue that the stacked QD has
different shape and different strain in and around the QD. These
effects affects the hole g-factor more than the electron and the
g-factor of the hole can even change sign \cite{nakaoka:235337}. The
X$^{-1}$ line of the blue QD has an exciton g-factor of 1.93.}. In
these plots, the QDs are in the gate bias regime where we see
X$^{1-}$ absorption in the blue QD and X$^0$ absorption in the red
QD. Based on the DT data shown and the unambiguous identification of
the charging states through charge sensing, we can with very high
confidence state that the excited states we probe in the DT
measurements of Fig. 2(b) involve one hole localized in the red QD
and two electrons. The DT signal disappears below a gate voltage of
150 mV, due to the fact that one of the optically generated
electrons become unstable and tunnels out in to the free electron
gas: in this regime, the red QD can be optically charged with a
hole.

The results shown in Figs.~2(b-c) indicate a different regime of
coherent coupling from those realized in previous studies on CQDs
\cite{krenner:076403}, where electron-electron exchange was found to
dominate the spin interactions. This can be seen most notably in the
ratio of the transition strengths of the anti-crossing and crossing
branches in Fig. 2(b) which here is about 1:1 as opposed to 1:3 in
previous studies (for comparison, a numerical calculation for that
regime is provided in the inset to Fig. 2(b)). Based on a simple
model that we present below, we attribute this qualitative difference
to an unusually strong intra-dot electron-hole exchange interaction.
We find that the following basis most transparently describes the DT
data\\

$ | 1 \rangle = e_{B\downarrow}^{\dagger} e_{B\uparrow}^{\dagger}
h_{R\Uparrow}^{\dagger} | 0 \rangle = ( \downarrow \uparrow ,
\Uparrow )$, \enspace \enspace \thinspace \thinspace $| 6 \rangle =
( \downarrow \uparrow , \Downarrow ),$

$ | 2 \rangle = e_{B\downarrow}^{\dagger} e_{R\downarrow}^{\dagger}
h_{R\Uparrow}^{\dagger} | 0 \rangle = ( \downarrow , B_+ )$,
\enspace  \enspace $ | 7 \rangle = ( \downarrow , B_- ),$

$ | 3 \rangle = e_{B\uparrow}^{\dagger} e_{R\downarrow}^{\dagger}
h_{R\Uparrow}^{\dagger} | 0 \rangle = ( \uparrow , B_+ )$, \enspace
\enspace $ | 8 \rangle = ( \uparrow , B_- ),$

$ | 4 \rangle = e_{B\downarrow}^{\dagger} e_{R\uparrow}^{\dagger}
h_{R\Uparrow}^{\dagger} | 0 \rangle = ( \downarrow , D_+ )$,
\enspace  \enspace $ | 9 \rangle = ( \downarrow , D_- ),$

$ | 5 \rangle = e_{B\uparrow}^{\dagger} e_{R\uparrow}^{\dagger}
h_{R\Uparrow}^{\dagger} | 0 \rangle = ( \uparrow , D_+ )$, \enspace
\enspace $ | 10 \rangle = ( \uparrow , D_- ),$ \\ \\ here,
$e_{i\sigma}^{\dagger}$ ($h_{i\sigma}^{\dagger}$) creates a
spin$-\sigma$ electron (hole) in the blue ($i=B$) or red ($i=R$) dot
($\Uparrow,\Downarrow$ refer to the hole pesudo-spin $J_z=\pm 3/2$).
The states $|1\rangle$ and $|6\rangle$ forms a spin singlet with
both electrons in the blue QD and the other states are compositions
of the one electron in the blue QD with a certain spin and a bright
$(B_\pm)$ or dark $(D_\pm)$ exciton in the red QD, where "$\pm$"
refers to right-hand/left-hand circular polarization. Considering
the states from $ | 1 \rangle $ to $ | 5 \rangle $, relevant for DT
measurements carried out using right-hand circularly polarized
laser, the Hamiltonian for this system and basis states is:

\[
\begin{pmatrix}
d_{i}\frac{V}{l} + U & 0 & t_e & -t_e & 0 \\
0 & E_{1} & 0 & 0 & 0 \\
t_e & 0 & E_{1} & 0 & 0 \\
-t_e & 0 & 0 & E_{2} & 0 \\
0 & 0 & 0 & 0 & E_{2}
\end{pmatrix}
\]\\ where $E_{1}=\frac{1}{2}\delta_0+d_{d}\frac{V}{l}$, 
$E_{2}=-\frac{1}{2}\delta_0+d_{d}\frac{V}{l}$, $V$
is the applied gate bias, $l$ is the distance between the back and
top gates, $d_{i}$ and $d_{d}$ are the sizes of effective indirect
and direct static dipoles and $U$ is the intra-dot Coulomb
interaction for two electrons in the blue QD measured with respect
to their inter-dot Coulomb interaction. We consider a Hamiltonian
that is extended in block diagonal form when we consider all 10
states as we can neglect the anisotropic electron-hole exchange
interaction \footnote{The X$^0$ of the red CQD1 has the unusual
attribute of an X-Y splitting that is less than the linewidth.}. 
Simulations has been carried out with the extended 10 state
Hamiltonian and an energy level diagram as a function of gate bias
$V$ for typical parameters is given in Fig. 2(d) with only states
$|1\rangle$ to $|5\rangle$ labeled. The calculated DT absorption
spectra are shown in Fig. 2(e) ($B=0$) and 2(f) ($B=0.4T$). The
applied magnetic field lifts the degeneracy of the different
electron-hole spin configuration. An important feature of this
Hamiltonian is the coupling of the state $|1\rangle$ with the states
$|3\rangle$ and $|4\rangle$ through the $t_e$ term, and as shown in
Fig. 2(d). This particular hybridization will brighten the dark
exciton $|4\rangle$ in a narrow gate bias regime around the
anti-crossing as seen in the PL data of Fig. 1(b) and (e) (Line
marked X$^D$). Note that the X$^D$ is brightest at a somewhat larger
gate bias than predicted.

By fitting our model to the PL and DT data, we determine the
isotropic part of electron-hole exchange interaction $\delta_0$ in
the red QD of CQD1 to be $230\mu eV$, while the single electron
coupling is estimated to be $t_e=137\mu eV$, indicating we are in a
new regime where the intra-dot electron-hole exchange interaction is
the dominant scale. The strength of the anti-crossing considered
($t^*=161\mu eV$ indicated in the inset of Fig. 2(e)) implies that
the exchange interaction is mainly mediated by \textit{one} electron
tunneling, while in the typical case of weak electron-hole exchange
the value is about $t^*=\sqrt{2}t_e$ arising from the possibility of
two electrons tunneling \cite{krenner:057402}. Here, stronger
electron-hole exchange ensures that effectively one electron spin is
available for tunneling.

\begin{figure}
\includegraphics[width=82mm]{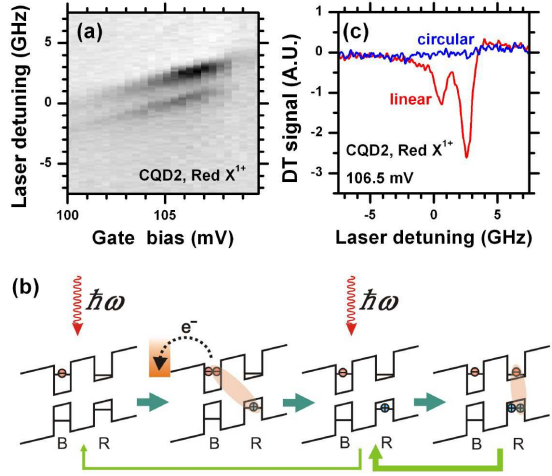}
\caption{\label{fig:hole} (a) DT measurements of X$^+$ with
excitation of only one laser. (b) Schematic description of the two-photon 
process resulting in scattering of the X$^{1+}$ transition with only one 
resonant laser (c) Polarization dependence of the
X$^+$ line, showing maximum absorption of linearly polarized light
and minimal for circularly polarized light.}
\end{figure}

We next investigate the gate bias regime  which allows us to switch
the electron-hole interaction off through an optical
double-resonance. This process is described in Fig. 3(b). Starting
with an electron in the blue dot, the excitation of an indirect
X$^0$ is followed by one of the electron tunneling to the
back-contact as a two-electron charge in the blue QD is unstable at
this gate bias, leaving a state with one electron in the blue and
one hole in the red QD. This is the ground state of the X$^{1+}$
transition which turns out to be at the same energy, which allows
scattering until the hole tunnels out towards the top gate resetting
the system to the initial configuration of one electron in the blue
dot. The relative brightness of the X$^{1+}$ to the X$^0$ in Fig.
1(e) indicates that the lifetime of the optically generated hole is
long compared to the lifetime of the exciton. Note that the indirect
X$^0$ is comparatively weaker in the shown range. The electron-hole
exchange interaction in the X$^{1+}$ configuration will vanish
because of the two holes in the red QD forming a singlet. Hence, the
two lines in Fig. 3(a) effectively correspond to
emission from an electronic singlet/triplet state formed by pure
indirect electron-electron exchange (A similar two-electron
coupling, but for the ground state was studied in
Ref.~\cite{Tureci-2006}). Figure 3(b) shows the DT signal at a
gate bias of 106.5 mV.  By using a linearly polarized resonant
laser, we find that the higher energy transmission dip that
corresponds to the three triplet states has an area close to three
times that of the lower energy one that corresponds to the singlet
state. The splitting between the dips is about 8.5 $\mu$eV. The
electron tunneling rate corresponding to the measured splitting is
found to be $t_e\approx140\mu eV$ which is close to the previously
calculated value of $t_e=137\mu eV$ for CQD1.

Further in Fig. 3(c), we present data showing a strong 
polarization dependence of the two photon absorption process in Fig. 3(a). 
The resonance is clearly visible
with linear polarized light, but with a circular polarized laser, it
is substantially reduced. The optical selection rules for the indirect X$^0$ 
and the direct X$^{1+}$ are orthogonal for circular polarization, thereby 
blocking either the optical charging of the hole or the scattering depending 
on the spin of the original electron in the blue QD.

In conclusion, we have demonstrated coherent coupling between two stacked QDs 
in the regime where the electron-hole exchange is the dominant interaction and 
leads to qualitative differences in the spectrum as compared to the case where 
the inter-dot indirect electron exchange process is the dominant scale. We were 
able to probe the opposite regime of nominally vanishing electron-hole exchange 
by varying the gate voltage and utilizing a double-resonance in absorption.  We 
expect that the optical charge sensing that we used to identify the charging 
states in our system is itself a very valuable tool for other applications in 
quantum information processing such as single spin measurement via spin-charge 
conversion \cite{PhysRevLett.70.1311,elzerman:430431}.

\begin{acknowledgments}
Y. Z. wishes to acknowledge funding from LGFG. This work was
supported by NCCR Quantum Photonics (NCCR QP), research instrument
of the Swiss National Science Foundation (SNSF).
\end{acknowledgments}

\newpage

\end{document}